\documentstyle[12pt]{article}

\def\noi{\noindent}
\def\lam{\lambda}

\def\del{\partial}
\def\nab{\nabla}
\def\til{\tilde}
\def\dis{\displaystyle}

\def\a{\rm a}
\def\b{\rm b}

\def\A{\rm A}
\def\B{\rm B}

\begin{document}

\begin{center}
{\Large\bf The equivalence theorem in the generalized gravity of $f(R)$-type
and canonical quantization}
\\[10mm]
\end{center}

\noi
\hspace*{1cm}\begin{minipage}{14.5cm}
Y. Ezawa$^1$ and Y.Ohkuwa$^2$\\

\noi
$^1$ Dept. of Physics, Ehime university, Matsuyama, 790-8577, Japan\\
$^2$ Section of Mathematical Science, Dept. of Social Medicine, Faculty 
of Medicine,\\
$\ $  University of Miyazaki, Kiyotake, Miyazaki, 889-1692, Japan\\

\noi
Email : ezawa@phys.sci.ehime-u.ac.jp, ohkuwa@med.miyazaki-u.ac.jp
\\

\noi
{\bf Abstract}\\[2mm]
We first review the equivalence theorem of the $f(R)$-type gravity to 
Einstein gravity with a scalar field by deriving it in a self-contained 
and pedagogical way.
Then we describe the problems of to what extent the equivalence holds.
Main problems are: (i) Is the surface term given by Gibbons and Hawking 
which is necessary in Einstein gravity also necessary in the $f(R)$-type 
gravity? (ii) Does the equivalence hold also in quantum theory? (iii) 
Which metric is physical, i.e., which metric should be identified with 
the observed one?
In this work, we clarify the problem (i) and review the problem (ii) in 
a canonical formalism which is the generalization of the Ostrogradski one.
We briefly comment on the problem (iii).

Some discussions are given on one of the results of (ii) concerning the 
general relativity in the non-commutative spacetime.
\end{minipage}

\section{Introduction}

Since the discovery of the accelerated expansion of the universe
\cite{SN,WMAP}, much attention has been attracted to the generalized 
gravity theories of the $f(R)$-type\cite{CDTT,SF,NO}.
Before the discovery, such theories have been interested in because of 
its theoretical advantages: 
The theory of the graviton is renormalizable\cite{UDW, Stelle}. 
It seems to be possible to avoid the initial singularity of the universe 
\cite{NT} which is the prediction of the theorem by Hawking\cite{Hawking}. 
And inflationary model without inflaton field is possible\cite{Staro}.

There is a well-known equivalence theorem between this type of theories 
and Einstein gravity with a scalar field\cite{Equiv}.
The theorem states that two types of theories related by a suitable 
conformal transformation are equivalent in the sense that the field 
equations of both theories lead to the same paths.
Many investigations have been devoted to this issue\cite{MS,SF}.
In this work, we first review classical aspects of the theorem in a 
self-contained way.
Next we solve the problem of the surface terms or the variational conditions.
The surface term is not necessary since we can impose the variational 
conditions at the time boundaries that the metric and its "time derivative" 
can be put to be vanishing.
This simplicity could be added to the advantages of $f(R)$-type gravity.
Quantum aspects of the theorem are then summarized when we quantize the 
theory canonically in the framework of the generalized Ostrogradski formalism 
which is a natural generalization to the system in a curved spacetime. 
The main result is that if the $f(R)$-type theory is quantized canonically, 
Einstein gravity with a scalar field has to be quantized non-canonically.

In section 2, the Lagrangian density and field equations for the $f(R)$-type 
gravity are summarized.
In section 3, the equivalence theorem is derived in a pedagogical way.
In section 4, the problems concerning the equivalence theorem are pointed out, 
especially to what extent the equivalence holds.
In section 5, the issue of surface term is clarified.
Section 6 is devoted to a description of the canonical formalism of the 
$f(R)$-type gravity in the Jordan and Einstein frame. 
Summary and discussions are given in section 7.
Summary of conformal transformations of geometrical quantities and the 
description of Ostrogradski transformation are given in the appendix.

\section{Generalized gravity of $f(R)$-type}

Generalized gravity of $f(R)$-type is one of the higher curvature 
gravity(HCG) theories in which the action is given by
$$
S=\int d^Dx{\cal L}=\int d^Dx\sqrt{-g}f(R).                        \eqno(2.1)
$$
The spacetime is taken to be $D$-dimensional. 
Here $g\equiv \det g_{\mu\nu}$ and $R$ is the $D$-dimensional scalar 
curvature.
Taking the variational conditions at the hypersurfaces $\Sigma_{t_{1}}$ 
and $\Sigma_{t_{2}}$ ($\Sigma_{t}$ is the hypersurface $t=constant$) as
$$
\delta g_{\mu\nu}=0\ \ \ {\rm and}\ \ \ \delta\dot{g}_{\mu\nu}=0, \eqno(2.2)
$$ 
field equations are derived by the variational principle as follows:
$$
-{\delta{\cal L}\over\delta g_{\mu\nu}(x)}
=\sqrt{-g}\Bigl[f'(R)R^{\mu\nu}-{1\over2}f(R)g^{\mu\nu}
-\nab^{\mu}\nab^{\nu}f'(R)+g^{\mu\nu}\Box f'(R)\Bigr]=0,           \eqno(2.3\a)
$$
or
$$
G_{\mu\nu}
={1\over f'(R)}\Bigl[\;{1\over2}\Bigl(f(R)-Rf'(R)\Bigr)g_{\mu\nu}
-(g_{\mu\nu}\Box-\nab_{\mu}\nab_{\nu})f'(R)\Bigr],                 \eqno(2.3\b)
$$
where a prime represents the differentiation with respect to $R$, 
$\nab_{\mu}$ the covariant derivative with respect to the metric
$g_{\mu\nu}$ and $G_{\mu\nu}\,$ is the $D$-dimensional Einstein tensor.
Equations (2.3a,b) are the 4-th order partial differential equations, 
so the above variational conditions are allowed.
Further discussions on this issue will be given in Section 5.

Here we comment on the dimensionality of $f(R)$.
Comparing the action $S$ with the Einstein-Hilbert one
$$
S_{E-H}={1\over 2\kappa_{D}^2}\int d^Dx\sqrt{-g}R,                 \eqno(2.4)
$$
where $\kappa_{D}\equiv\sqrt{8\pi G_{D}}$ with $G_{D}$ the $D$-dimensional 
gravitational constant, we obtain the dimension of $f'(R)$ to be equal 
to that of $\kappa_{D}^{\;-2}$, so that
$$
[\,f'(R)\,]=[\,\kappa_{D}^{\;-2}\,]=[\,L^{2-D}\,].                 \eqno(2.5)
$$

It is well known that this type of theory is transformed to Einstein 
gravity with a scalar field by a conformal transformation, which is 
usually referred to as equivalence theorem.
We will review and clarify the content of the theorem.

\section{Equivalence theorem}

The theorem concerns with the conformal transformation
$$
\til{g}_{\mu\nu}\equiv \Omega^2g_{\mu\nu}.                         \eqno(3.1)
$$
In terms of the transformed Einstein tensor, field equations (2.3b) are written 
as


$$
\hspace*{-5.15mm}\begin{array}{ll}
\til{G}_{\mu\nu}&\dis \!\!\!={1\over f'(R)}\nab_{\mu}\nab_{\nu}f'(R)
-(d-1)\nab_{\mu}\nab_{\nu}(\ln \Omega)
-g_{\mu\nu}\Bigl[{1\over f'(R)}\Box f'(R)-(d-1)\Box(\ln \Omega)\Bigr]
\\[5mm]
&\dis +(d-1)\del_{\mu}(\ln \Omega)\del_{\nu}(\ln \Omega)
+g_{\mu\nu}\Bigl[{f(R)-Rf'(R)\over 2f'(R)}
+{(d-1)(d-2)\over 2}\del_{\lam}(\ln \Omega)\del^{\lam}(\ln \Omega)\Bigr],
\end{array}                                                        \eqno(3.2)
$$
where we put $D\equiv 1+d$ (i.e. $d$ is the dimension of the space).
Eqs.(3.2) are the field equations after the conformal transformation.
If they are the equations for Einstein gravity with a scalar field, 2nd order 
derivatives on the right hand side should vanish.
From this requirement, $\Omega$ is determined to be
$$
\Omega^2=\left[2\kappa_{D}^{\;2}f'(R)\right]^{2/(d-1)}.            \eqno(3.3)
$$
The coefficient of $f'(R)$ in the square bracket, which can be any 
constant, was chosen to be $2\kappa_{D}^{\;2}$ in order to make 
$\Omega$ to be dimensionless and equal to unity for Einstein gravity.
So, (3.1) takes the following form
$$
\til{g}_{\mu\nu}
=\left[2\kappa_{D}^{\;2}f'(R)\right]^{2/(d-1)}g_{\mu\nu}.          \eqno(3.4)
$$
Scalar field is defined as
$$
\kappa_{D}\,\til{\phi}\equiv \sqrt{d(d-1)}\ln \Omega
=\sqrt{d/(d-1)}\ln[2\kappa_{D}^2 f'(R)],                           \eqno(3.5\a)
$$
or
$$
f'(R)
={1\over 2\kappa_{D}^{\;2}}\exp{\Bigl(\sqrt{(d-1)/d}\,\kappa_{D}\,
\til{\phi}\,\Bigr)},\ \ 
\ln \Omega={1\over\sqrt{d(d-1)}}\,\kappa_{D}\,\til{\phi}.          \eqno(3.5\b)
$$
The coefficient of $\ln \Omega$, or equivalently $\ln[2\kappa_{D}^2 f'(R)]$, 
in (3.5a) was chosen for the right-hand side of (3.2) to take the usual 
form of scalar field source.
Solving (3.5) for $R$, we denote the solution as
$$
R=r(\til{\phi}).                                                   \eqno(3.6)
$$
In terms of $\til{\phi}$, (3.2) takes the following form
$$
\til{G}_{\mu\nu}=\kappa_{D}^{\;2}\left[\del_{\mu}\til{\phi}\,
\del_{\nu}\til{\phi}+\til{g}_{\mu\nu}\Bigl(-{1\over2}\del_{\lam}
\til{\phi}\,\til{\del}^{\lam}\til{\phi}-V(\til{\phi})\Bigr)\right], \eqno(3.7)
$$
where 
$\til{\del}^{\lam}\til{\phi}\equiv \til{g}^{\lam\rho}\del_{\rho}\til{\phi}$ 
and
$$
V(\til{\phi})\equiv 
-f\left(r(\til{\phi})\right)\exp{\Bigl(-{d+1\over \sqrt{d(d-1)}}\kappa_{D}\,
\til{\phi}\,\Bigr)}
+{1\over 2\kappa_{D}^{\;2}}\,r(\til{\phi})\exp{\Bigl(-{2\over
\sqrt{d(d-1)}}\kappa_{D}\,\til{\phi}\,\Bigr)}.
                                                                   \eqno(3.8)
$$
Field equation for the scalar field is obtained by taking the trace 
of (3.2) as
$$
\widetilde{\Box}{\til{\phi}}
=-{\kappa_{D}\over \sqrt{d(d-1)}}
\exp{\Bigl(-{d+1\over \sqrt{d(d-1)}}\kappa_{D}{\til{\phi}}\Bigr)}
\Bigl[(d+1)f(r(\til{\phi}))-\kappa_{D}^{\;-2}\,r(\til{\phi})\exp{
\Bigl(\sqrt{(d-1)/d}\,\kappa_{D}\til{\phi}\Bigr)}\Bigr].           \eqno(3.9)
$$
Equations (3.7) and (3.9) are obtained also by the variational 
principle with the following Lagrangian density:


$$
\til{\cal L}=\til{\cal L}_{G}+\til{\cal L}_{\til{\phi}},           \eqno(3.10)
$$
where
$$
\til{\cal L}_{G}={1\over 16\pi G_{D}}\sqrt{-\til{g}}\til{R}
,\ \ \ \ \ 
\til{\cal L}_{\til{\phi}}
=\sqrt{-\til{g}}\Bigl[-{1\over2}\del_{\lam}\til{\phi}\,
\til{\del}^{\lam}\til{\phi}-V(\til{\phi})\Bigr].                   \eqno(3.11)
$$
Here
$$
\sqrt{-\til{g}}
=\left[2\kappa_{D}^{\;2}f'(R)\right]^{(d+1)/(d-1)}\,\sqrt{-g},     \eqno(3.12)
$$
and
$$
\til{R}=\left[2\kappa_{D}^{\;2}f'(R)\right]^{-2/(d-1)}\Bigl[R-{2d\over d-1}
\Bigl({1\over f'(R)}\Box f'(R)-{1\over2}{1\over f'(R)^2}
\del_{\lam}f'(R)\del^{\lam}f'(R)\Bigr)\Bigr].                      \eqno(3.13)
$$
$\til{\cal L}_{\til{\phi}}$ is given by terms in the parenthesis 
multiplying $\til{g}_{\mu\nu}$ in (3.7) and $V(\til{\phi})$ is given by (3.8).
It is noted that this Lagrangian density $\til{\cal L}$ is not equal 
to the Lagrangian density ${\cal L}$ in (2.1) which, in terms of the 
transformed variables $\til{g}_{\mu\nu}$ and $\til{\phi}$, is expressed as
$$
{\cal L}=\sqrt{-\til{g}}f\left(r(\til{\phi})\right)\exp{\Bigl(
-{d+1\over\sqrt{d(d-1)}}\,\kappa_{D}\til{\phi}\Bigr)}.
$$

Thus from the field equations (2.3b) for the $f(R)$-type gravity, field 
equations for $\til{g}_{\mu\nu}$ with the source of the scalar field 
and the field equation for the scalar field are derived.
So the equivalence seems to be shown.
However, eqs.(2.3b) are 10 4-th order differential equations for 10 
component $g_{\mu\nu}$, so that, 
to obtain a unique set of solutions, 40 initial conditions seem to 
be required.
On the other hand eqs.(3.7) are 10 2nd order differential equations for 
10 component $\til{g}_{\mu\nu}$, only 20 initial conditions are required 
to have a set of unique solution.
Similarly, eq.(3.9) requires only 2 initial conditions.
Therefore equivalence does not hold if the initial conditions are taken 
into account.
This apparent breakdown comes from the fact that the 40 initial conditions 
are not independent, which is easily seen in canonical formalism (see 
section 5).

The above result that the variational equations of both theories 
coincide is usually stated as “HCG described by the Lagrangian 
density ${\cal L}$ is equivalent to Einstein gravity with a scalar field 
described by the Lagrangian density $\til{\cal L}\,$" and is 
referred to as the equivalence theorem.
Note, however, that the variational equations hold on the paths 
that make the action stationary.
Ref.\cite{MS} is recommended as a good review on the equivalence theorem.
For recent investigations, see Ref.\cite{FN} and references cited in 
these references.
We use the following usual terminology on this issaue:
$$
\left\{\begin{array}{lcl}
{\rm descriptions\ with}\ {\cal L}&：&{\rm descriptions\ in\ the\ 
Jordan\ frame}\\
{\rm descriptions\ with}\ \til{\cal L}&:&{\rm descriptions\ in\ the\ 
Einstein\ frame}
\end{array}\right.
$$

\section{Problems}

We have seen that the equivalence of the two theories hold at 
least on the classical paths which can be determined by the 
variational principle.
However, there would be problems on the other kinds of equivalence.
In order to examine these problems, we note the following:\\[1mm]
$
\hspace*{1cm}\left[\begin{array}{l}
1.\ {\rm The\ theories\ are\ not\ conformally\ invariant}.\\
2.\ {\rm The\ physical\ metric\ is\ identified\ with\ the\ 
one\ determined\ from\ observations}.
\end{array}\right.
$
\\[2mm]
Unsettled problems include the following:\\
(I) To what extent the equivalence would hold?\\
(I-1) In the Einstein frame, it is well known that the surface term 
given by Gibbons and Hawking (GH  term)\cite{GH} is necessary.
It is often argued that, from the equivalence point of view, surface 
term is necessary also in the Jordan frame\cite{EqST}.
However, this equivalence is not taken for granted, but should be 
examined carefully.
The examination is given in the next section.\\
(I-2) Would the equivalence hold also in quantum theory? 
If the equivalence holds in the canonical quantum theories, fundamental
Poisson brackets should be equivalent.
That is, the fundamental Poisson brackets in one frame should be derived 
from those of the other frame.\\
(II) Which metric is physical in the sense that should be identified 
with the observed one?\\
This problem has been investigated from various aspects\cite{MS}.
If the metric in the Einstein frame is physical\cite{Phy}, HCG has 
no essential meaning and it appears by the choice of unphysical frame.
If the metric in the Jordan frame is physical, the equivalence theorem 
states that the metric in this frame has one more scalar degrees of 
freedom which could be observed as non-transverse-traceless polarization 
of gravitational waves\cite{AMdeA} in future observations. 
Furthermore, equivalence theorem states that, instead of treating the 
complicated Jordan frame, we can use the simpler and familiar Einstein 
frame for calculation.
However, for comparison with observations, the results should be 
expressed in the words of Jordan frame. 
It should be noted only one of the metrics is physical.
In the following, assuming that the metric in the Jordan frame is 
physical, we restrict ourselves to the description of problem (I).

\section{Surface terms}

\subsection{General considerations}

We first consider discrete systems whose Lagrangians contain the time 
derivatives of the generalized coordinates $q^i$ up to the $n$-th order 
$q^{i(n)}$.
If the $n$-th order derivatives are contained non-linearly the equations 
of motion are $2n$-th order differential equations.
Then $2n$ conditions are necessary to determine the solution uniquely.
These conditions can be given by $2n$ initial conditions or $n$ 
boundary conditions at two times, $t_{1}$ and $t_{2}$.
The latter conditions can be taken to be the values of the generalized 
coordinates themselves and their time derivatives up to the $(n-1)$-th order.
Then we can take the variational conditions (boundary conditions) as
$$
\delta q^{i(k)}(t_{1})=\delta q^{i(k)}(t_{2})=0,\ \ \ 
(k=0,1,\cdots, n-1).                                               \eqno(5.1)
$$
Therefore no boundary terms are necessary.

On the other hand, if the $n$-th order derivatives are contained 
linearly, equations of motion are at most $(2n-1)$-th order 
differential equations.
Then at least one condition in (5.1) does not hold generally.
Therefore special solutions are required to satisfy all the conditions 
in (5.1) and to eliminate generally the corresponding variations at 
the boundaries, boundary terms are necessary.
In other words, in order that the equations of motion and the variational 
conditions are compatible, boundary terms are required.

For continuous systems, or fields, we can proceed similarly, i.e. 
if the Lagrangian contains the highest order derivatives linearly, 
surface terms are required to eliminate some of the variations of 
derivatives at the boundaries.

\subsection{$f(R)$-type gravity}

In this theory, the Lagrangian density contains the components of the 
metric, the generalized coordinates, and their derivatives up to the 
second order in a non-linear way.
So from the general considerations above, no surface terms are necessary.
Concrete situations are as follows.

The variational principle leads to the field equations which are 4-th order 
differential equations as noted above, so that 40 conditions are formally 
required to decide the solution for the metric uniquely, although they are 
not independent.
These conditions can be taken to be the initial functions of the components 
of the metric $g_{\mu\nu}$ itself and their derivatives up to the 3rd order, 
or $g_{\mu\nu}$ and their first order derivatives at 2 times $t=t_{1}$ and 
$t=t_{2}$.
The latter conditions correspond to the variational conditions at the time 
boundaries.
That is, at 2 time boundaries $t=t_{1}$ and $t=t_{2}$, variational conditions 
are taken as  $\delta g_{\mu\nu}=0$ and $\delta\dot{g}_{\mu\nu}=0$ as are 
given by (2.2).
In fact the Lagrangian density contains up to the 2nd order derivatives 
non-linearly, no surface term is necessary.

\subsection{Einstein gravity with a scalar field}

In this theory, the gravity theory is the Einstein one and if we start from 
the Lagrangian density $\til{\cal L}$, (3.10), whose gravitational part 
$\til{\cal L}_{G}$ contains the second order derivatives of the metric 
linearly, surface term e.g. the GH term, is necessary from the above 
considerations.
Some arguments exist that if we require the equivalence also in the 
boundary terms, surface term is necessary also in the $f(R)$-type 
gravity\cite{EqST}.
This is not the case. This equivalence should be examined carefully. 
The situation can be seen by examining the variation.
If the theory is obtained from the $f(R)$-type theory by the conformal 
transformation, $\til{g}_{\mu\nu}=\left[2\kappa_{D}^{\;2}f'(R)\right]^{2/(d-1)}
g_{\mu\nu}$. 
So, if we express the variation of this quantity and $\til{\phi}$ in terms of 
the variations in the Jordan frame, we have the following relations:
$$
\left\{\begin{array}{cl}
\delta\til{g}_{\mu\nu}&
=\dis\left[2\kappa_{D}^{\;2}f'(R)\right]^{2/(d-1)}\delta g_{\mu\nu}
+{4\kappa_{D}^{\;2}\over d-1}\left[2\kappa_{D}^{\;2}f'(R)\right]^
{-(d-3)/(d-1)}g_{\mu\nu}\,\delta f'(R)
\\[5mm]
\delta\til{\phi}&\dis 
=\kappa_{D}^{\;-1}\sqrt{d/(d-1)}\,{1\over f'(R)}\,\delta f'(R),
\end{array}\right.                                                 \eqno(5.2)
$$
where
$$
\delta f'(R)={\del f'\over \del g_{\alpha\beta}}\delta g_{\alpha\beta}
+{\del f'\over \del(\del_{\lam}g_{\alpha\beta})}
\delta(\del_{\lam}g_{\alpha\beta})
+{\del f'\over\del(\del_{\lam}\del_{\rho}g_{\alpha\beta})}
\delta(\del_{\lam}\del_{\rho}g_{\alpha\beta}).                     \eqno(5.3)
$$
Therefore, if both sets of the variational conditions 
$$
\delta\til{g}_{\mu\nu}=\delta\til{\phi}=0,                         \eqno(5.4)
$$
which are usually taken for $\til{\cal L}$ and the boundary conditions (2.2), $\delta g_{\mu\nu}
=\delta\dot{g}_{\mu\nu}=0$, are imposed, we have
$$
\delta\ddot{g}_{\mu\nu}=0,                                         \eqno(5.5)
$$
at the boundary.
However, this is not generally possible, but would require specific 
solutions as noted above.
That is, the variational conditions, which require the GH term in the 
Einstein gravity with a scalar field, are different from those 
in the $f(R)$-type theory.
To compare the surface terms, the variational conditions have to be 
carefully treated.

The above situation is related to the fact that the conformal transformation 
is not the transformation of the generalized coordinates, $g_{\mu\nu}$, but 
the transformation depending on the 2nd order derivatives of them.
Comparison of the surface terms is made as follows.
When $\til{\cal L}$ is expressed in terms of the metric in the Jordan frame, 
$g_{\mu\nu}$, it is written as follows:
$$
\til{\cal L}
={\cal L}-\del_{\lam}\Bigl({2d\over d-1}\sqrt{-g}\,\del^{\lam}f'(R)\Bigr).
                                                                   \eqno(5.6)
$$
Since ${\cal L}$ requires no surface term when the variational condition 
(2.2) are taken, the second term on the right-hand side is the surface term 
which is different from the GH term. 
This is an example that surface terms depend on the boundary conditions.

\section{Canonical formalism}

The canonical formalism belongs to classical physics.
However, most quantum theory is obtained by canonical quantization which 
requires that commutation relations among the fundamental quantities are 
proportinal to the corresponding Poisson brackets, e.g. for one dimensional 
system
$$
[\hat{q},\hat{p}]=i\hbar\{q,p\}_{PB},
$$
where a hat represents an operator.
It is noted that one of the proportional factor $i$ assures the hermiticity 
of observables and the other $\hbar$ adjusts the dimensionality, a very 
natural proportional factors.

Canonical quantum theories are very successful and only well-known failure 
is the theory of graviton in general relativity.
On the other hand, the canonical quantum theory of gravitons in $f(R)$-type 
gravity is known to be renormalizable\cite{UDW,Stelle}.
This suggests a possibility that the equivalence theorem would be violated 
in quantum theory.
The violation might come from the fact that classical equivalence means the 
equivalence along the classical paths.
While, the Poisson brackets require derivatives in all directions in the 
phase space.
The laws of usual canonical quantum theory describe the dynamics of matter 
and radiation which have duality of waves and particles assured by 
experiments.
On the other hand, gravity describes the dynamics of spacetime.
However, no nature of spacetime similar to the duality has been observed.
Investigation of quantum gravity arises from various motivations.
For example, since the gravity mediates interactions of elementary particles, 
it would be natural that the gravity is also described quantum mechanically.
A preferable possibility that fundamental laws of nature would take forms of 
quantum theory is also one of them.
The canonical quantum theory would be the first candidate for quantum gravity.
Therefore a canonical formalism of gravity is very important.
In this section results on a canonical formalism, a generalization of 
the Ostrogradski formalism, are reviewed.
In the following, we use a unit for which $2\kappa_{D}^{\;2}=1$.

\subsection{Canonical formalism in the Einstein frame}

We adopt the ADM method for the gravitational field\cite{ADM}, so the 
procedure is well known.

\subsubsection{Gravitational field}

The spacetime is supposed to be constructed from the hypersurfaces 
$\Sigma_{t}$ with $t=constant$ (foliation of spacetime).
The dynamics of the spacetime determines the evolution of the hypersurface.
So the generalized coordinates are the metric of the $d$-dimensional 
hypersurface  $\til{h}_{ij}({\bf x},t)$.

Since $\til{R}$ contains 2nd order time derivatives linearly, we first 
make a partial integration to transform the Lagrangian density 
of the gravitational part in (3.11) to the following GH form:
$$
\til{\cal L}_{h}
=\sqrt{\til{h}}\,\til{N}\,\bigl[\til{K}_{ij}\til{K}^{ij}
-\til{K}^2+\til{\bf R}\;\bigr],                                    \eqno(6.1)
$$
where $\til{h}\equiv\det \til{h}_{ij}$ and $\til{N}$ is the lapse 
function and $\til{K}$ is the trace of the extrinsic curvature 
$\til{K}_{ij}(\til{K}\equiv \til{h}^{ij}\til{K}_{ij})$ and 
$\til{\bf R}$ is the scalar curvature constructed from $\til{h}_{ij}$.
The extrinsic curvature $\til{K}_{ij}$ with respect to $\til{h}_{ij}$ 
is defined as
$$
\til{K}_{ij}\equiv 
{1\over2}\til{N}^{-1}\left(\del_{0}\til{h}_{ij}
                     -\til{N}_{i;j}-\til{N}_{j;i}\right),          \eqno(6.2)
$$
where $\til{N}_{i}$ is the shift vector.
A semicolon ; represents a covariant derivative with respect to 
$\til{h}_{ij}$.

Canonical formalism is obtained by the Legendre transformation as usual.
The momenta $\til{\pi}^{ij}$ canonically conjugate to $\til{h}_{ij}$
are defined as
$$
\til{\pi}^{ij}\equiv
{\del\til{\cal L}_{h}\over\del(\del_{0}\til{h}_{ij})}
=\sqrt{\til{h}}\left[\til{K}^{ij}-\til{h}^{ij}\til{K}\right]       \eqno(6.3)
$$

Solving (6.3) for $\til{K}_{ij}$, we have
$$
\til{K}^{ij}={1\over\sqrt{\til{h}}}\left[\til{\pi}^{ij}
            -{1\over d-1}\til{h}^{ij}\til{\pi}\right]
\ \ \ {\rm and}\ \ \ 
\til{K}=-{\til{\pi}\over (d-1)\sqrt{\til{h}}}.                     \eqno(6.4)
$$
Hamiltonian density is given as
$$
\begin{array}{lcl}
\til{\cal H}_{h}&=&\til{\pi}^{ij}\dot{\til{h}}_{ij}-\til{{\cal L}_{h}}
\\[3mm]
&=&\dis 
\til{N}\,\Bigl[\,G_{ijkl}\til{\pi}^{ij}\til{\pi}^{kl}
   -\sqrt{\til{h}}{\bf \til{R}}\Bigr]
   +2(\til{\pi}^{ij}\til{N}_{i})_{;j}-2\til{\pi}^{ij}_{\ ;j}\til{N}_{i}.
\end{array}                                                        \eqno(6.5)
$$
where
$$
G_{ijkl}\equiv
{1\over2\sqrt{h}}\Bigl(\,\til{h}^{ik}\til{h}^{jl}
+\til{h}^{il}\til{h}^{jk}-{2\over d-1}\til{h}^{ij}\til{h}^{kl}\,\Bigr),
                                                                   \eqno(6.6)
$$
is sometimes referred to as supermetric.
In deriving (6.5), we used the expression for $\til{\cal L}_{h}$, expressed 
in terms of canonical variables, as follows
$$
\til{\cal L}_{h}
={\til{N}\over\sqrt{\til{h}}}\left[\,\til{\pi}^{ij}\til{\pi}_{ij}
-{1\over d-1}\til{\pi}^2+\til{h}\til{\bf R}\;\right].              \eqno(6.7)
$$

\subsubsection{Scalar field}

The generalized coordinate is $\til{\phi}({\bf x},t)$.
Momenta canonically conjugate to $\til{\phi}$ is defined as usual by


$$
\til{\pi}({\bf x},t)\equiv
{\del\til{\cal L}_{\phi}\over\del(\del_{0}\til{\phi}({\bf x},t))}
=-\sqrt{\til{g}\;}\,\til{g}^{0\mu}\del_{\mu}\til{\phi}
=\til{N}^{-1}\sqrt{\til{h}}\left[\del_{0}\til{\phi}
-\til{N}^i\del_{i}\til{\phi}\right],                               \eqno(6.8\a)
$$
so
$$
\del_{0}\til{\phi}={\til{N}\over \sqrt{\til{h}}}
\left[\,\til{\pi}+\til{N}^{-1}\sqrt{\til{h}}\til{N}^i\del_{i}\til{\phi}\;
\right]
={\til{N}\over\sqrt{\til{h}}}\,\til{\pi}
+\til{N}^i\del_{i}\til{\phi}.                                      \eqno(6.8\b)
$$
In terms of canonical variables, $\til{\cal L}_{\phi}$ is expressed 
as follows
$$
\til{\cal L}_{\phi}=\til{N}\left[{1\over 2\sqrt{\til{h}}}\,\til{\pi}^2
-{1\over 2}\sqrt{\til{h}}\,\til{h}^{ij}\del_{i}\til{\phi}\,\del_{j}\til{\phi}
-V(\til{\phi})\right].
$$
Using this, we have the following expression for the Hamiltonian density
$$
\begin{array}{lcl}
\til{\cal H}_{\phi}&=&\til{\pi}\dot{\til{\phi}}-\til{\cal L}_{\phi}
\\[3mm]
&=&\dis {\til{N}\over2\sqrt{\til{h}}}\,\til{\pi}^2
+\til{N}^i\del_{i}\til{\phi}\;\til{\pi}
+{1\over2}\til{N}\sqrt{\til{h}}\,\til{h}^{ij}\del_{i}\til{\phi}\,
\del_{j}\til{\phi}+V(\til{\phi}).
\end{array}                                                        \eqno(6.9)
$$

\subsubsection{Fundamental Poisson brackets}

Nonvanishing fundamental Poisson brackets in the Einstein frame are given as
$$
\{\til{h}_{ij}({\bf x},t),\til{\pi}^{kl}({\bf y},t)\}_{PB}
=\delta_{(ij)}^{kl}\delta({\bf x}-{\bf y})\ \ \ {\rm and}\ \ \ 
\{\til{\phi}({\bf x},t),\til{\pi}({\bf y},t)\}_{PB}
=\delta({\bf x}-{\bf y}),                                          \eqno(6.10)
$$
where $(ij)$ expresses the symmetrization and not the symmetric part.

\subsection{Canonical formalism in the Jordan frame}

There are several canonical formalisms for generalized gravity theories 
in the Jordan frame.
Among them formalism given by Buchbinder and Lyakhovich\cite{BL} is logically 
very simple.
However, concrete calculation is somewhat cumbersome partly due to 
arbitrariness although it allows a wide application.
In addition, the Hamiltonian is generally transformed under the transformation 
of generalized coordinates that does not depend on time explicitly.
Here we use the formalism which is a generalization of the well-known 
one given by Ostrogradski\cite{EIOWYY}.
For comparison of typical formalisms, see \cite{DSY}.

\subsubsection{Generalized coordinates}


\noi
In this frame, we also use the foliation of the spacetime.
Since the $f(R)$-type gravity is a higher-derivative theory, we follow 
the modified Ostrogradski formalism in which the time derivatives in 
the Ostrogradski formalism is replaced by Lie derivatives along the 
timelike normal to the hypersurface $\Sigma_{t}$ in the ADM formalism
\cite{EIOWYY}.
So the generalized coordinates are
$$
h_{ij}({\bf x},t)\ \ \ {\rm and}\ \ \ 
K_{ij}({\bf x},t)
={1\over2}{\cal L}_{n}h_{ij}({\bf x},t)\equiv Q_{ij}.              \eqno(6.11)
$$
Here contravariant and covariant components of the normal $n$ are 
given as follows:
$$
n^{\mu}=N^{-1}(1,-N^i)\ \ \ {\rm and}\ \ \ n_{\mu}=N(-1,0,0,0).    \eqno(6.12)
$$

\subsubsection{Conjugate momenta}

Denoting the momenta canonically conjugate to these generalized 
coordinates as $\pi^{ij}$ and $\Pi^{ij}$ respectively, we have from 
the modified Ostrogradski transformation


$$
\left\{\begin{array}{l}
\pi^{ij}=-\sqrt{h}\left[f'(R)Q^{ij}+h^{ij}f''(R){\cal L}_{n}R\,\right]
\\[3mm]
\Pi^{ij}=2\sqrt{h}f'(R)h^{ij}.
\end{array}\right.                                                 \eqno(6.13)
$$
From (6.13), it is seen that $\Pi^{ij}$ has only the trace part, so 
it is expressed as
$$
\Pi^{ij}={1\over d}\Pi h^{ij}\ \ \ {\rm and}\ \ \ 
\Pi=2d\sqrt{h}f'(R).                                               \eqno(6.14)
$$
From the second equation, we have
$$
f'(R)={\Pi\over 2d\sqrt{h}}\ \ \ {\rm or}\ \ \ 
R=f'^{-1}(\Pi/2d\sqrt{h})\equiv \psi(\Pi/2d\sqrt{h}).              \eqno(6.15)
$$
Correspondingly, it is also seen from (6.13) that the traceless part of 
$Q_{ij}$ is related to that of $\pi^{ij}$, and we have
$$
Q^{ij}=-{2\over P}\pi^{\dagger ij}+{1\over d}h^{ij}Q,              \eqno(6.16)
$$
where
$$ P\equiv {\Pi\over d},                                           \eqno(6.17)
$$
and
$$
\pi^{\dagger ij}\equiv \pi^{ij}-{1\over d}h^{ij}\pi                \eqno(6.18)
$$
is the traceless part. A dagger is used to represent the traceless part.
$(Q,P)$ is one of the canonical pairs.
In terms of these variables, the scalar curvature is expressed as follows
$$
R=2h^{ij}{\cal L}_{n}Q_{ij}+Q^2-3Q_{ij}Q^{ij}+{\bf R}-2\Delta(\ln N)
                                                                   \eqno(6.19)
$$

\subsubsection{Hamiltonian density}

In the modified Ostrogradski formalism, Hamiltonian density is 
defined as
$$
{\cal H}\equiv \pi^{ij}\dot{h}_{ij}+\Pi^{ij}\dot{Q}_{ij}-{\cal L}. \eqno(6.20)
$$
Using
$$
{\cal L}_{n}Q_{ij}=N^{-1}(\del_{0}Q_{ij}-N^kQ_{ij;k}-N^k_{\ ;i}Q_{kj}
-N^k_{\ ;j}Q_{ik}-N^{-1}\del_{i}N\del_{j}N)                        \eqno(6.21)
$$
and eqs. (6.14)$-$(6.19), we have an explicit expression for ${\cal H}$ 
as follows:
$$
\hspace*{-5mm}\begin{array}{ll}
\hspace*{-1mm}{\cal H}&
=\dis N\Bigl[{2\over P}\pi^{\dagger ij}\pi^{\dagger}_{\ ij}
+{2\over d}Q\pi+{1\over2}P\psi(P/2\sqrt{h})-{d-3\over 2d}Q^2P
+{1\over2}{\bf R}P+\Delta P-\sqrt{h}f\left(\psi(P/2\sqrt{h})\right)\Bigr]
\\[5mm]
&\dis  +N^k\Bigl[2\pi^{\dagger\ \ ;j}_{\ kj}-{2\over d}\pi_{:k}
+P\del_{i}Q-{2\over d}(QP)_{;k}\Bigr]
\\[5mm]
&\dis +\left[-2N_{j}\pi^{ij}+N^j(\pi+QP)+\del^iNP-NP^{;i}\right]_{;i}
\end{array}                                                        \eqno(6.22)
$$

\subsubsection{Fundamental Poisson brackets}

Non-vanishing fundamental Poisson brackets are the following:


$$
\{h_{ij}({\bf x},t),\pi^{kl}({\bf y},t)\}_{PB}
=\delta^i_{(k}\delta^j_{l)}\delta({\bf x}-{\bf y}),               \eqno(6.23\a)
$$
and
$$
\{Q_{ij}({\bf x},t),\Pi^{kl}({\bf y},t)\}_{PB}
=\delta^i_{(k}\delta^j_{l)}\delta({\bf x}-{\bf y}).               \eqno(6.23\b)
$$

\subsubsection{Wheeler-DeWitt equation}

A primary application of the canonical formalism is the Wheeler-DeWitt
(WDW) equation.
Before writing down the WDW equation, we make a canonical 
transformation
$$
(Q,P)\rightarrow (\bar{Q},\bar{P})\equiv (P,-Q),                   \eqno(6.24)
$$
which removes the negative powers of the momentum $P$.
The resulting Hamiltonian is expressed as follows:
$$
{\cal H}=N{\cal H}_{0}+N^k{\cal H}_{k}+{\rm divergent\ term},      \eqno(6.25)
$$
where
$$
\left\{\begin{array}{ll}
\dis{\cal H}_{0}&\dis 
={2\over Q}\,\pi^{\dagger ij}\pi^{\dagger}_{\ ij}-{2\over d}P\pi
+{1\over2}Q\psi(Q/2\sqrt{h})-{d-3\over 2d}QP^2-{1\over 2}{\bf R}Q
\\[5mm]
&\dis -\sqrt{h}\,f\left(\psi(Q/2\sqrt{h})\right)+\Delta Q,
\\[7mm]
{\cal H}_{k}&\dis 
=2\pi^{\dagger\ \ ;j}_{\;kj}-{2\over d}\pi_{;k}-QP_{;k}+{2\over d}(QP)_{;k}.
\end{array}\right.                                                 \eqno(6.26)
$$
The WDW equation is written as
$$
\hat{\cal H}_{0}\Psi=0,                                            \eqno(6.27)
$$
where $\hat{\cal H}_{0}$ is obtained from ${\cal H}_{0}$ by replacing 
$\pi^{ij}$ and $P$ with $-i\del/\del h_{ij}$ and $-i\del/\del Q$, 
respectively.
However, in order to apply (6.27) to the observed universe after 
compactification, we first carry out the dimensional reduction and then 
we should take into account the cosmological principle.
Such procedures were done using the formalism of Buchbinder and 
Lyakhovich which, although is generally different from the one 
described above, is very similar in the case of gravity\cite{EKKSY}.
It was shown by the semiclassical approximation method that the 
internal space could be stabilized.

\subsection{Compatibility of the two sets of fundamental Poisson brackets}

\subsubsection{Compatibility conditions}

The canonical variables in the Einstein frame can be expressed in terms 
of those in the Jordan frame.
So we can calculate the left hand sides of (6.10) using (6.23a,b).
The compatibility conditions are that the results are the right hand 
sides of (6.10), i.e. the following relations should be satisfied:
$$
\begin{array}{ll}
\{\til{h}_{ij}({\bf x},t),\til{\pi}^{kl}({\bf y},t)\}_{PB}&
=\dis\sum_{m,n}\int d^d{\bf z}\Biggl[\left\{
{\del\til{h}_{ij}({\bf x},t)\over\del h_{mn}({\bf z},t)}
{\del\til{\pi}^{kl}({\bf y},t)\over\del p^{mn}({\bf z},t)}-
{\del\til{\pi}^{kl}({\bf y},t)\over\del h_{mn}({\bf z},t)}
{\del\til{h}_{ij}({\bf x},t)\over p^{mn}({\bf z},t)}\right\}\\[5mm]
&\hspace{1.5cm}+\dis \left\{
{\del\til{h}_{ij}({\bf x},t)\over\del Q_{mn}({\bf z},t)}
{\del\til{\pi}^{kl}({\bf y},t)\over\del\Pi^{mn}({\bf z},t)}-
{\del\til{\pi}^{kl}({\bf y},t)\over\del Q_{mn}({\bf z},t)}
{\del\til{h}_{ij}({\bf x},t)\over \Pi^{mn}({\bf z},t)}\right\}\Biggr]
\\[5mm]
&=\delta^i_{(k}\delta^j_{l)}\delta ({\bf x}-{\bf y}),
\end{array}                                                        \eqno(6.28)
$$
and
$$
\begin{array}{ll}
\{\til{\phi}({\bf x},t),\til{\pi}({\bf y},t)\}_{PB}&
=\dis\sum_{m,n}\int d^d{\bf z}
\Biggl[\left\{{\del\til{\phi}({\bf x},t)\over\del h_{mn}({\bf z},t)}
{\del\til{\pi}({\bf y},t)\over\del p^{mn}({\bf z},t)}-
{\del\til{\pi}({\bf y},t)\over\del h_{mn}({\bf z},t)}
{\del\til{\phi}({\bf x},t)\over p^{mn}({\bf z},t)}\right\}\\[5mm]
&\hspace{1.6cm}\dis +\left\{{\del\til{\phi}({\bf x},t)\over
\del Q_{mn}({\bf z},t)}
{\del\til{\pi}({\bf y},t)\over\del \Pi^{mn}({\bf z},t)}-
{\del\til{\pi}({\bf y},t)\over\del Q_{mn}({\bf z},t)}
{\del\til{\phi}({\bf x},t)\over \Pi^{mn}({\bf z},t)}\right\}\Biggr]
\\[5mm]
&=\delta ({\bf x}-{\bf y}).
\end{array}                                                        \eqno(6.29)
$$
Other fundamental Poisson brackets should vanish.
These conditions may lead to some restrictions on $f(R)$.

\subsubsection{Expression of the conformal transformation in terms 
of canonical variables}

Using (3.4),(6.2),(6,3) and (6.8a,b), we obtain the following form 
of the conformal transformation expressing the canonical variables 
in the Einstein frame in terms of those in the Jordan frame:


$$
\hspace*{-3mm}\left\{\begin{array}{ccl}
\til{h}_{ij}&=&\dis f'(R)^{2/(d-1)}h_{ij}
               =\left(P/2\sqrt{h}\right)^{2/(d-1)}h_{ij}
\\[5mm]
\til{\phi}&
=&\dis \sqrt{d/(d-1)}\,\ln\left(P/2\sqrt{h})\right)
\\[5mm]
\til{\pi}&
=&\dis \sqrt{d/2(d-1)}N^{-1}\left[
\del_{0}P-P(NQ+N^i_{\ ;i})+N^iP_{;i}\right]
\\[5mm]
\til{N}&=&\dis\left(P/2\sqrt{h}\right)^{1/(d-1)}\,N,\ \ \ \til{N}^i=N^i
\\[5mm]
\til{\pi}^{ij}
&=&\dis\left(P/2\sqrt{h}\right)^{(d-3)/(d-1)}\sqrt{h}\biggl[-{2\over P}
\pi^{\dagger ij}
+h^{ij}\Bigl\{{1\over d}Q-(NP)^{-1}\left(\del_{0}P-N^kP_{;k}\right)
-N^{-1}N^k_{\ ;k}\Bigr\}\biggr].
\end{array}\right.                                                 \eqno(6.30)
$$

\subsubsection{Calculation of the Poisson brackets}

It may seem that the calculations are carried out easily.
However, the evaluations of the brackets involving the time derivatives 
of the momenta are difficult.
It is noted that it is impossible to use the field equations. 
Since, in that case, changes of variables are restricted to those along 
the paths of motions, which does not fit to Poisson brackets which use 
changes in any direction.
Nevertheless, we can show, using (6.30), that assumption that all of 
the equations (6.10),(6.17),(6.18) leads to contradiction\cite{EIOWYY}.
In other words, two frames are not related by a canonical transformation.

Therefore, in the framework of the canonical formalism used here, we 
cannot quantize the theory canonically in both frames.
That is, if the $f(R)$-type theory is quantized canonically, corresponding 
Einstein gravity with a scalar field has to be quantized non-canonically, 
e.g. in the non-commutative geometric way.

\section{Summary and discussions}

In this work, we reviewed the equivalence theorem in the $f(R)$-type 
gravity by deriving it in a pedagogical and self-contained way.
Equivalence of this theory with Einstein gravity with a scalar field, 
related by a conformal transformation, holds on the classical paths.
Strictly speaking, description in the physical frame is equivalent to 
the description in the unphysical frame, since only one frame is physical.
If the description in the unphysical frame is simpler, calculations 
could be done in the frame.

Concerning the surface term in the $f(R)$-type gravity, it is not necessary 
in the Jordan frame.
Necessity of the surface term in the Einstein frame comes from the structure 
of the Lagrangian density that it contains the 2nd order derivatives linearly.
A concrete example of the surface term is obtained that shows the dependence 
of it on the variational conditions.
The usual variational conditions in the Einstein frame leads to the GH term.
On the other hand, if the variational conditions are taken as in the Jordan 
frame, the surface term is different and is given in (5.6).

In the canonical formalism, the conformal transformation is not a canonical 
one.
So the fundamental Poisson brackets are not equivalent in the sense that the 
sets of fundamental Poisson brackets in both frames are not compatible.
Thus if the theory is quantized canonically in the Jordan frame, quantization 
in the Einstein frame has to be non-canonical, e.g. in the non-commutative 
geometric way\cite{Kempf}.
It is pointed out that similar situation occurs in the inflation model in 
multidimensional Einstein gravity\cite{ESWY}.
In this model, the $n$-dimensional internal space continues to shrink during 
inflation and loses its gravitational potential energy which is transferred 
to the inflating space. 
The potential energy behaves as $a_{I}^{-(n-2)}$, which is expected by the 
Gauss law in $n$-dimensional space, so that the shrinkage of the internal 
space leads classically to the collapse of the internal space similar to the 
situation in the case of atoms. 
However if $n>3$, the canonical quantum theory cannot prevent the collapse 
 of the internal space contrary to the case of atoms, so that non-canonical 
quantum theory is required.
Recently, in the noncommutative geometric multidimensional cosmology, it is 
shown that stabilization of the internal space is possible\cite{KJS}. 
This suggests that in the multidimensional $f(R)$-type gravity, 
extra-dimensional space would be stable.
This result is in conformity with that obtained by the semiclassical 
approximation to the WDW equation noted above.

Thus, considering the renormalizability of the graviton theory, stabilization 
of the internal space in the semiclassical approximation to WDW equation, it 
is plausible that $f(R)$-type gravity can be quantized canonically in the 
Jordan frame.
In addition, similar stabilization is possible in noncommutative geometric 
way, so quantization in the Einstein frame is non-canonical.
\\[8mm]

\appendix

\noi
{\Large\bf Appendix}

\section{Conformal transformations of geometrical quantities}

\noi
We consider a conformal transformation given as
$$
\til{g}_{\mu\nu}\equiv \Omega^2g_{\mu\nu}.                         \eqno(\A.1)
$$
Transformations of geometrical quantities are given as follows.\\[2mm]
{\bf Christoffel symbols}
$$
\til{\Gamma}^{\lam}_{\mu\nu}=\Gamma^{\lam}_{\mu\nu}
+\delta^{\lam}_{\mu}\,\del_{\nu}(\ln \Omega)
+\delta^{\lam}_{\nu}\,\del_{\mu}(\ln \Omega)
-g_{\mu\nu}\,\del^{\lam}(\ln \Omega).                              \eqno(\A.2)
$$
{\bf Covariant derivatives}\\[1mm]
For a scalar field, we have
$$
\til{\nab}_{\mu}\til{\nab}_{\nu}\phi
=\nab_{\mu}\nab_{\nu}\phi-\left[\del_{\mu}(\ln \Omega)\del_{\nu}\phi
+\del_{\nu}(\ln \Omega)\del_{\mu}\phi
-g_{\mu\nu}\del^{\lam}(\ln \Omega)\del_{\lam}\phi\right]          \eqno(\A.3\a)
$$
or
$$
\widetilde{\Box}\phi
\equiv \til{g}^{\mu\nu}\til{\Delta}_{\mu}\Delta_{\nu}\til{\phi}
=\Omega^{-2}\left[\Box\phi
              +(D-2)\del^{\lam}(\ln \Omega)\del_{\lam}\phi\right] \eqno(\A.3\b)
$$
{\bf Ricci tensor}
$$
\til{R}_{\mu\nu}=R_{\mu\nu}-(D-2)[\nab_{\mu}\nab_{\nu}(\ln \Omega)
-\del_{\mu}(\ln\Omega)\del_{\nu}(\Omega)]-g_{\mu\nu}[\Box(\ln \Omega)
+(D-2)\del_{\lam}(\ln \Omega)\del^{\lam}(\ln \Omega)]              \eqno(\A.4)
$$
{\bf scalar curvature}
$$
\til{R}=\Omega^{-2}\left[R-2(D-1)\Box(\ln \Omega)
-(D-1)(D-2)\del_{\lam}(\ln \Omega)\del^{\lam}(\ln \Omega)\right]   \eqno(\A.5)
$$
{\bf Einstein tensor}
$$
\til{G}_{\mu\nu}=G_{\mu\nu}-(D-2)\Bigl[\nab_{\mu}\nab_{\nu}(\ln \Omega)
-g_{\mu\nu}\Box(\ln \Omega)-\del_{\mu}(\ln \Omega)\del_{\nu}(\ln \Omega)
-{D-3\over2}g_{\mu\nu}\del_{\lam}\del^{\lam}(\ln \Omega)\Bigr]     \eqno(\A.6)
$$

\section{Canonical formalism by Ostrogradski}

Here we introduce the description of Ostrogradski's canonical formalism 
given by T. Kimura and R. Sugano\cite{K-S} adding a simple example, 
however restricting only to the regular case.

\subsection{A simple example}

We begin with a simple example of a system with one degree of freedom 
and the Lagrangian of the system depends on the generalized coordinate $q$ 
and its time derivatives up to the second order:
$$
L=L(q,\dot{q},\ddot{q}).                                           \eqno(\B.1)
$$
The action $S$ is given as
$$
S[q]=\int_{t_{1}}^{t_{2}}L(q,\dot{q},\ddot{q})\,dt.                \eqno(\B.2)
$$

\subsubsection{Variational principle}

Variation of this action, (B.2), is as follows:
$$
\delta S\equiv S[q+\delta q]-S[q]
=\int_{t_{1}}^{t_{2}}\delta Ldt
=\int_{t_{1}}^{t_{2}}\Bigl[{\del L\over \del q}\delta q
+{\del L\over \del\dot{q}}\delta\dot{q}+{\del L\over \del\ddot{q}}
\delta\ddot{q}\Bigr]dt.                                            \eqno(\B.3)
$$
Making partial integrations, we have
$$
\delta S
=\Bigl[\Bigl\{{\del L\over \del\dot{q}}
-{d\over dt}\Bigl({\del L\over\del\ddot{q}}\Bigr)\Bigr\}\delta q
+{\del L\over \del\ddot{q}}\delta\dot{q}\Bigr]_{t_{1}}^{t_{2}}
+\int_{t_{1}}^{t_{2}}\Bigl[{\del L\over \del q}
-{d\over dt}\Bigl({\del L\over\del\dot{q}}\Bigr)
+{d^2\over dt^2}\Bigl({\del L\over \del\ddot{q}}\Bigr)\Bigr]\delta qdt.
                                                                   \eqno(\B.4)
$$
In applying the variational principle, we need boundary conditions 
for the integration at $t=t_{1}$ and $t=t_{2}$.
We adopt the following boundary conditions, i.e., 
$$
\delta q=\delta\dot{q}=0\ \ \ 
{\rm at}\ \ \ t=t_{1}\ {\rm and}\ t=t_{2}.                         \eqno(\B.5)
$$
Variational principle requires that the action is stationary for 
arbitrary $\delta q$ except for the boudaries.
Then we have the following equation of motion:
$$
{\del L\over\del q}-{d\over dt}\Bigl({\del L\over\del\dot{q}}\Bigr)
+{d^2\over dt^2}\Bigl({\del L\over\del\ddot{q}}\Bigr)=0.           \eqno(\B.6)
$$
This is the generalized Euler-Lagrange equation which is the 4-th 
order differential equation unless the time derivative of the second 
order is included linearly in the Lagrangian.
Therefore the boundary conditions (B.5) are allowed.

\subsubsection{Ostrogradski transformation}

Now in order to transform to the canonical formalism, in which the 
equations of motion are 1st order differential equations, we introduce 
new generalized coordinates to lower the order of the time derivatives 
as follows:
$$
q_{0}\equiv q,\ \ \ q_{1}\equiv \dot{q}.                           \eqno(\B.7)
$$
The momenta canonically conjugate to these generalized coordinates 
are defined to be the coefficients of the variations of these 
generalized coordinates in the boundary terms in eq.(B.4).
Explicitly, they are expressed as
$$
p_{0}\equiv {\del L\over \del\dot{q}}
            -{d\over dt}\Bigl({\del L\over \del\ddot{q}}\Bigr),\ \ \ 
p_{1}\equiv {\del L\over \del\ddot{q}}.                            \eqno(\B.8)
$$
When the Lagrangian does not depend on the second order time derivatives, 
this definition reduces to the usual one.
Thus (B.8) is the generalization of the usual ones.
Using these momenta, we define the Hamiltonian as follows:
$$
H\equiv \sum_{s=0}^{1}p_{s}\dot{q}_{s}-L(q_{0},q_{1};\dot{q}_{1}). \eqno(\B.9)
$$
By considering the change, we can see that this Hamiltonian is a 
function of the new generalized coordinates and the momenta canonically 
conjugate to them.
The change of the Hamiltonian is the following:
$$
dH= \sum_{s=0}^1 dp_{s}\dot{q}_{s}+p_{0}d\dot{q}_{0}+p_{1}d\dot{q}_{1}
  -\Bigl({\del L\over \del q_{0}}dq_{0}+{\del L\over \del q_{1}}dq_{1}
  +{\del L\over \del\dot{q}_{1}}d\dot{q}_{1}\Bigr).
$$
Here the following relations hold:
$$
d\dot{q}_{0}=dq_{1},\ \ \ 
{\del L\over \del\dot{q}_{1}}={\del L\over \del\ddot{q}}=p_{1}.
$$
Therefore we have
$$
dH=\sum_{s=0}^1 dp_{s}\dot{q}_{s}+p_{0}dq_{1}-\Bigl({\del L\over 
\del q_{0}}dq_{0}+{\del L\over \del q_{1}}dq_{1}\Bigr)
$$
which shows that $H$ is a function of only $q_{s}$ and $p_{s}$ and 
not $\dot{q}_{s}$.
Now we change the description in terms of a set 
$({\rm Lagrangian},q,\dot{q},\ddot{q})$ to the description in terms of 
a set $({\rm Hamiltonian}, q_{s},p_{s})$.
This change will be referred to as {\bf Ostrogradski transformation}.

\subsubsection{Canonical equations of motion}

Since $H$ depends on only $q_{s}$ and $p_{s}$, its variation is expressed as
$$
\delta H
=\sum_{s=0}^1\Bigl[{\del H\over \del q_{s}}\delta q_{s}
+{\del H\over \del p_{s}}\delta p_{s}\Bigr].                       \eqno(\B.10)
$$
On the other hand, from the definition of H, (B.9), the variation is 
expressed as
$$
\delta H
=\sum_{s=0}^1[\delta p_{s}\dot{q}_{s}+p_{s}\delta\dot{q}_{s}]
-\delta L.                                                         \eqno(\B.11)
$$
The first term on the right hand side is rewritten as
$$
\sum_{s=0}^1\Bigl[\dot{q}_{s}\delta p_{s}+{d\over dt}(p_{s}\delta q_{s})
-\dot{p}_{s}\delta q_{s}\Bigr].
$$
On the second term, we have
$$
\delta L
=\Bigl[{\del L\over \del q}-{d\over dt}\Bigl({\del L\over \del\dot{q}}\Bigr)
+{d^2\over dt^2}\Bigl({\del L\over \del\ddot{q}}\Bigr)\Bigr]\delta q
+{d\over dt}\Bigl(\sum_{s=0}^1p_{s}\delta q_{s}\Bigr).
$$
When the Euler-Lagrange equation is satisfied, the quantity in the curly 
bracket on the right hand side vanishes. Then the variation, (B.11), 
takes the following form:
$$
\delta H
=\sum_{s=0}^1[\dot{q}_{s}\delta p_{s}-\dot{p}_{s}\delta q_{s}].    \eqno(\B.12)
$$
Therefore from (B.10) and (B.12), we have
$$
\dot{q}_{s}={\del H\over \del p_{s}},\ \ \ 
\dot{p}_{s}=-{\del H\over \del q_{s}},\ \ \ (s=0,1).               \eqno(\B.13)
$$
These equations are the canonical equations of motion.
We can show that the Hamiltonian is invariant under the transformation 
of the generalizeed coordinate $q\rightarrow Q\equiv f(q)$.

\subsection{Generalization to a system with $N$ degrees of freedom}

Here we consider a system with $N$ degrees of freedom whose generalized 
coordinates are denoted as $q^i,\ (i=1,2,\cdots,N)$.
Its Lagrangian, L, is assumed to depend on the time derivatives of 
these coordinates up to $n$-th order:
$$
L=L(q^i,\dot{q}^i,\cdots,q^{i(n)}).                                \eqno(\B.14)
$$
We could generalize further such that the orders of the highest time 
derivatives are different for each $i$, i.e., $n\rightarrow n_{i}$.
However, we do not make this generalization, as it does not require 
essentially new elements and only complicate equations.
We could proceed pararelly to those in the case of example above.
Instead, we start from a slightly general variation of the action, 
i.e. boundaries of integration are also varied, which leads to the 
definition of the Hamiltonian.

\subsubsection{Generalized variation}

We will denote the variation of the action noted above as 
$\delta^{*} S$ which is expresed as follows:
$$
\delta^{*}S\equiv \int_{t_{1}+\delta t_{1}}^{t_{2}+\delta t_{2}}
L\left(q^i+\delta^{*} q^i,\dot{q}^i+\delta^{*} \dot{q}^i,\ldots,
q^{i(n)}+\delta^{*} q^{i(n)}\right)dt
-\int_{t_{1}}^{t_{2}}L\left(q^i,\dot{q}^i,\ldots,q^{i(n)}\right)dt,\eqno(\B.15)
$$
where
$$
\delta^{*} q^i\equiv (q+\delta q)^i(t+\delta t)-q^i(t).            \eqno(\B.16)
$$
Rewriting the right hand side, we have
$$
\delta^{*}q^i=\left[(q+\delta q)^i(t+\delta t)-q^i(t+\delta t)\right]
           +\left[q^i(t+\delta t)-q^i(t)\right].                  \eqno(\B.17)
$$
The quantity in the first curly bracket on the right hand side 
expresses the variation of only the coordinates which is used in 
the usual variational principle and will be denoted as $\delta q^i$. 
Therefore we have
$$
\delta^{*}q^i=\delta q^i+\dot{q}^i\delta t.                       \eqno(\B.18)
$$
The first integral on the right hand side of (B.15) are rewritten 
as follows:
$$
\int_{t_{1}+\delta t_{1}}^{t_{2}+\delta t_{2}}
=\int_{t_{1}+\delta t_{1}}^{t_{1}}+\int_{t_{1}}^{t_{2}}
+\int_{t_{2}}^{t+\delta t_{2}}
=\int_{t_{1}}^{t_{2}}+\int_{t_{2}}^{t+\delta t_{2}}
-\int_{t_{1}}^{t_{1}+\delta t_{1}}.
$$
Taking $\delta t_{1}$ and $\delta t_{2}$ to be small, we can 
approximate, e.g.
$$
\int_{t_{1}}^{t_{1}+\delta t_{1}}L\,dt\approx [L\delta t]_{t=t_{1}}.
$$
Therfore we can approximate as
$$
\delta^{*}S=\Bigl[L\delta t\Bigr]_{t_{1}}^{t_{2}}+\delta S.        \eqno(\B.19)
$$
Varying the time $t$ only near 2 boundaries of integration, we have
$$
\delta S\equiv \int_{t_{1}}^{t_{2}}
L\left(q^i+\delta q^i,\dot{q}^i+\delta \dot{q}^i,\ldots,
q^{i(n)}+\delta q^{i(n)}\right)dt
-\int_{t_{1}}^{t_{2}}L\left(q^i,\dot{q}^i,\ldots,q^{i(n)}\right)dt.\eqno(\B.20)
$$
This is the usual variation used in the variational principle, 
so is the generalization of (B.4).
Evaluation of the right hand side, which leads to the generalized 
Euler-Lagrange equations, is carried out in the next subsetion. 
Before procdeeding, the following notation is introduced for simplicity:
$$
D\equiv {d\over dt}.                                              \eqno(\B.21)
$$

\subsubsection{Generalized variation}

Expressing the Lagrangian as $L(D^sq^i),\ (i=1,\cdots,N;\;s=0,1,\cdots,n)$,
we have for the variation of the action
$$
\delta S=\int_{t_{1}}^{t_{2}}\delta L\,dt,\ \ \ 
\delta L
=\sum_{i=1}^N\sum_{s=0}^n{\del L\over \del(D^sq^i)}\delta(D^sq^i),\ \ 
\delta(D^sq^i)=D^s(\delta q^i).                                    \eqno(\B.22)
$$
In order to obtain the Euler-Lagrange equation, repeated integrations 
by parts are required.
The first steps are the following:
$$
\begin{array}{ll}
\hspace*{-3mm}\dis {\del L\over \del(D^sq^i)}\delta(D^sq^i)&
\dis ={\del L\over \del(D^sq^i)}D^s(\delta q^i)
=D\Bigl[{\del L\over \del(D^sq^i)}D^{s-1}(\delta q^i)\Bigr]
-D\Bigl\{{\del L\over \del(D^sq^i)}\Bigr\}D^{s-1}(\delta q^i)
\\[5mm]
&=\cdots
\\[5mm]
&\dis 
=D\Bigl[\sum_{r=0}^{s-1}(-1)^rD^r\Bigl\{{\del L\over \del(D^sq^i)}
\Bigr\}\delta(D^{s-r-1}q^i\Bigr\}\Bigr]
+(-1)^sD^s\Bigl\{{\del L\over \del(D^sq^i)}\Bigr\}\delta q^i.
\end{array}                                                        \eqno(\B.23)
$$
Summing for $s(\geq 1)$, we have
$$
\begin{array}{l}
\dis \sum_{s=1}^n{\del L\over \del(D^sq^i)}\delta(D^sq^i)
\\[5mm]
\dis =D\Bigl[\sum_{s=1}^n\sum_{r=0}^{s-1}(-1)^rD^r\Bigl\{
{\del L\over \del(D^sq^i)}\Bigr\}\delta(D^{s-r-1}q^i)\Bigr]
+\sum_{s=1}^n(-1)^sD^s\Bigl\{{\del L\over\del(D^sq^i)}\Bigr\}\delta q^i
\\[5mm]
\dis =D\Bigl[\sum_{a=1}^n\delta(D^{a-1}q^i)\sum_{s=a}^n(-1)^{s-a}
D^{s-a}\Bigl\{{\del L\over \del(D^sq^i)}\Bigr\}\Bigr]
+\sum_{s=1}^n(-1)^sD^s\Bigl\{{\del L\over \del(D^sq^i)}\Bigr\}\delta q^i.
\end{array}                                                        \eqno(\B.24)
$$
Thus, we have
$$
\delta S=\Bigl[\delta F\Bigr]_{t_{1}}^{t_{2}}
+\int_{t_{1}}^{t_{2}}\sum_{i=1}^N\Bigl[\sum_{s=0}^n(-1)^sD^s\Bigl\{
{\del L\over\del(D^sq^i)}\Bigr\}\delta q^i\Bigr]dt,               \eqno(\B.25)
$$                                                                
where
$$
\delta F\equiv \sum_{i=1}^N\Bigl[\sum_{a=1}^n\delta(D^{a-1}q^i)
\sum_{s=a}^n(-1)^{s-a}D^{s-a}
\Bigl\{{\del L\over\del(D^sq^i)}\Bigr\}\Bigr].                    \eqno(\B.26)
$$
Therefore, we have
$$
\delta^{*}S=\Bigl[L\delta t+\delta F\Bigr]_{t_{1}}^{t_{2}}
+\int_{t_{1}}^{t_{2}}\sum_{i=1}^N\Bigl[\Bigl\{\sum_{s=0}^n(-1)^sD^s
\Bigl({\del L\over\del(D^sq^i)}\Bigr)\Bigr\}\delta q^i\Bigr]dt.   \eqno(\B.27)
$$ 

\subsubsection{Variational principle}

As noted above, $\delta S$, expressed by (B.25), is the variation of 
the action used in the variational principle.
Thus if we adopt the boundary conditions
$$
\delta(D^{a-1}q^i)=0,\ \ i=1,\cdots,N;\;a=1.\cdots,n\ \ \ 
{\rm at}\ \ t=t_{1}\ \ {\rm and}\ \ t=t_{2},                       \eqno(\B.28)
$$
we have from (B.27) or (B.25), the generalized Euler-Lagrange equations
(equations of motion) as
$$
{\del L\over \del q^i}
+\sum_{s=1}^n(-1)^sD^s\Bigl[{\del L\over \del(D^sq^i)}\Bigr]=0,
\ \ \ (i=1,\cdots,N).                                              \eqno(\B.29)
$$
These equations are $2n$-th order differential equations, unless 
$n$-th order derivatives are not contained linearly in the Lagrangian.
Therefore the boundary conditions (B.28) are allowed as in the case 
of the example.

\subsubsection{Ostrogradski transformation}

In order to transform to the canonical formalism, we first define 
the new generalized coordinates to lower the order of time derivatives 
of the generalized coordinates as follows:
$$
q^i_{s}\equiv D^sq^i\ \ \ (i=1, \cdots,N;\ s=0,1,\cdots,n-1).     \eqno(\B.30)
$$
Momenta canonically conjugate to these coordinates, $p^s_{i}$, 
are defined to be the coefficients of the variations of these 
coordinates, $\delta^{*}q^i_{s}$, in the boundary terms which is 
the quantities in the square bracket in the last line in eq.(B.24) 
or $\delta F$.
Explicitly they are given as
$$
p^s_{i}\equiv 
\sum_{r=s+1}^n(-1)^{r-s-1}D^{r-s-1}\Bigl\{{\del L\over \del(D^rq^i)}\Bigr\},
\ \ \ (i=1,\cdots,N;\;s=0,1,\cdots,n-2)                          \eqno(\B.31\a)
$$
except for $p^{n-1}_{i}$ which is defined as
$$
p_{i}^{n-1}\equiv 
{\del L\over \del\dot{q}^i_{n-1}}={\del L\over \del(D^nq^i)}.    \eqno(\B.31\b)
$$
Sometimes the canonical momenta are defined by the recursion formulae
$$
p^{s-1}_{i}={\del L\over \del(D^sq^i)}-Dp^s_{i},                  \eqno(\B.31c)
$$
which are derived from (B.31a) with (B.31b).
Using these new generalized coordinates and the momenta, we have
$$
\delta F\equiv \sum_{i=1}^N\sum_{s=0}^{n-1}\delta q_{s}^i\;p^s_{i}
=\sum_{i=1}^N\sum_{s=0}^{n-1}\left(\delta^{*}q_{s}^i
-\dot{q}_{s}^i\delta t\right)\,p^s_{i}.
$$
Therefore we have for the boundary terms in (B.27)
$$
\Bigl[\delta L+\delta F\Bigr]_{t_{1}}^{t_{2}}
=\Bigl[\Bigl(L-\sum_{i=1}^N\sum_{s=0}^{n-1}\dot{q}_{s}^i\;p^s_{i}\Bigr)
\delta t+\sum_{i=1}^N\sum_{s=0}^{n-1}\delta^{*}q_{s}^i\;p^s_{i}
\Bigr]_{t_{1}}^{t_{2}}.                                           \eqno(\B.32)
$$
Hamiltonian is defined to be the coefficient$\times(-1)$ of $\delta t$ 
on the right hand side, i.e.
$$
H\equiv 
\sum_{i=1}^N\sum_{s=0}^{n-1}p_{i}^s\dot{q}^i_{s}
-L(q^i_{0},q^i_{1},\cdots,q^i_{n-1};\dot{q}^i_{n-1}).              \eqno(\B.33)
$$
As in the case of the example above, the Ostrogradski transformation 
reduces to the Legendre transformation if the hightest order time 
derivatives are the first order
\footnote{The dimension of the velocity phase space, where the 
coordinates are the generalized coordinates and their time derivatives 
up to the n-th order, is $N(n+1)$. On the other hand, the dimension 
of the phase space, which has as coordinates the new generalized 
coordinates and the momenta canonically conjugate to them, is $2Nn$. 
So the dimensions of these two spaces are different. 
However, if the variational principle is imposed, we have constraints, 
$q^i_{s+1}=\del H/\del p^s_{i}$, coming from the definition of the new 
generalized coordinates, $\dot{q}^i_{s}=q^i_{s+1}$ and the canonical 
equations of motion, $\dot{q}^i_{s}=\del H/\del p^s_{i}$. 
The number of the constraints is $N(n-1)$. Thus the dimension of the 
subspace which satisfys the equation of motion (and could be referred 
to as physical subspace) is the same as that of velocity phase space. 
For $n=1$, there is no constraint of this type and the dimensions of 
the velocity phase space and the phase space are the same, which is 
the well known fact with respect to the Legendre transformation which 
is defined definitely without reference to the variational principle.}.
By examining the change of the Hamiltonian, we can show that it 
depends only on the generalized coordinates and the momenta canonically 
conjugate to them as in the case of the example.
The proof can be done as follows, quite pararelly as in the case of 
the example.
The change of the Hamiltonian is given by
$$
dH
=\sum_{i=1}^N\sum_{s=0}^{n-1}
(dp^s_{i}\,\dot{q}^i_{s}+p_{i}^s\,d\dot{q}^i_{s})-dL.             \eqno(\B.34)
$$
Using relations
$$
\left\{\begin{array}{l}
\dis \sum_{i=1}^N\sum_{s=0}^{n-1}p^s_{i}\,d\dot{q}^i_{s}
=\sum_{i=1}^N\Bigl[\sum_{s=0}^{n-2}p^s_{i}\,dq^i_{s+1}
 +p_{i}^{n-1}\,d\dot{q}^i_{n-1}\Bigr],
\\[5mm]
\dis dL
=\sum_{i=1}^N\Bigl[\sum_{s=0}^{n-1}{\del L\over\del q^i_s}\,dq_{s}^i
+{\del L\over\del\dot q^i_{n-1}}\,d\dot{q}^i_{n-1}\Bigr]
=\sum_{i=1}^N\Bigl[\sum_{s=0}^{n-1}{\del L\over \del q^i_{s}}\,dq^i_{s}
+p_{i}^{n-1}\,d\dot{q}^i_{n-1}\Bigr],
\end{array}\right.
$$
we have
$$
dH
=\sum_{i=1}^N\Bigl[\sum_{s=0}^{n-1}\dot{q}^i_{s}\,dp^s_{i}
 +\sum_{s=0}^{n-2}p^s_{i}\,dq^i_{s+1}
 -\sum_{s=0}^{n-1}{\del L\over \del q^i_{s}}\,dq^i_{s}\Bigr].      \eqno(\B.35)
$$
Therefore the Hamiltonian is a function of only $q^i_{s}$ and $p_{i}^s$.

\subsubsection{Canonical equations of motion}

Since the Hamiltonian is a function of only $q^i_{s}$ and $p_{i}^s$ 
as shown above, its variation is expressed as
$$
\delta H
=\sum_{i=1}^N\Bigl[\sum_{s=0}^{n-1}\Bigl(
           {\del H\over \del q^i_{s}}\delta q^i_{s}
           +{\del H\over \del p^s_{i}}\delta p^s_{i}\Bigr)\Bigr]. \eqno(\B.36)
$$
On the other hand, from the definition of the Hamiltonian, (B.34), 
we have
$$
\delta H
=\sum_{i=1}^N\Bigl[\sum_{s=0}^{n-1}\Bigl(\delta p_{i}^s\dot{q}^i_{s}
+p_{i}^s\delta\dot{q}_{s}^i\Bigr)\Bigr]-\delta L
=\sum_{i=1}^N\Bigl[\sum_{s=0}^{n-1}\Bigl\{\dot{q}_{s}^i\delta p_{i}^s
-\dot{p}_{i}^s\delta q_{s}^i+D\Bigl(p_{i}^s\delta q^i_{s}\Bigr)\Bigr\}
\Bigr]-\delta L.
$$
Here from the last line of eq.(B.24) and the definitions of 
$q^i_{s}$ and $p_{i}^s$, $\delta L$ is expressed as
$$
\delta L
=\sum_{i=1}^N\sum_{s=0}^{n-1}D(p_{i}^s\delta q^i_{s})
+\sum_{i=0}^N(E\!\!-\!\!L)_{i}\delta q^i.                          \eqno(\B.37)
$$
Here $(E\!\!-\!\!L)_{i}$'s are the left hand side of (B.29), so 
the generalized Euler-Lagrange equation is written as 
$(E\!\!-\!\!L)_{i}=0,\ \ (i=1,\cdots,N)$.
Thus, if the equations of motion are satisfied $\delta H$ takes the 
following form
$$
\delta H
=\sum_{i=1}^N\sum_{s=0}^{n-1}\Bigl[\dot{q}_{s}^i\delta p_{i}^s
-\dot{p}_{i}^s\delta q_{s}^i\Bigr].                                \eqno(\B.38)
$$
Therefore we have from (B.35) and (B.37) the following equations
$$
\dot{q}^i_{s}={\del H\over \del p^s_{i}},\ \ \ \ \ 
\dot{p}^s_{i}=-{\del H\over \del q^i_{s}}.                         \eqno(\B.39)
$$
That is to say, if the Euler-Lagrange equations are satisfied, 
canonical equations of motion are also satisfied.

Finally we comment on the generalization of the formalism described 
above.
In the generalized theory of gravity, e.g., $f(R)$-type one, modification 
of the formalism is necessary from the viewpoint of general relativity, 
if the ADM variables are used as noted in the text.
The situations are the following: The scalar curvature $R$ depends on 
the time derivatives of the lapse function and shift vector. 
So if we apply the method of Ostrogradski directly, these variables 
should be determined by solving the equations of motion.
This contradicts the general relativity which requires that these 
variables should be chosen arbitrarily, since the choice of them 
corresponds to the choice of the coordinate system.
The modification which replaces the time derivatives of the generalized 
coordinates by their Lie derivatives along the normal to the hypersurface 
of constant time function which is the time direction in the coordinate 
system with vanishing shift vector\cite{EIOWYY}.
This seem to be a natural and least modification.

\end{document}